\documentclass[useAMS,usenatbib,letterpaper]{mn2e} 
\usepackage{times}
\usepackage{amssymb}
\usepackage{graphicx}
\usepackage{multirow} 

\newcommand{\es}{1ES 1959+650}
\newcommand{\wco}{W Comae}
\newcommand{\rgb}{RGB J0710+591}
\newcommand{\coone}{$^{12}$CO($1-0$)}
\newcommand{\cotwo}{$^{12}$CO($2-1$)}
\newcommand{\alphaunit}{M$_\odot$ (K km s$^{-1}$ pc$^2$)$^{-1}$}
\newcommand{\msun}{M$_\odot$}
\newcommand{\lco}{$L'_{\rm CO}$}
\newcommand{\lcounit}{K km s$^{-1}$ pc$^2$}
\newcommand{\mhmol}{$M_{\rm H_2}$}
\newcommand\ion[2]{#1$\;${\sevensize#2}}%

\input{journal.def}

\title[On the low molecular gas content of BL Lac objects.]{A search of CO emission lines in blazars: the low molecular gas content of BL Lac objects compared to quasars.\thanks{Based on observations carried out with the IRAM 30m Telescope. IRAM is supported by INSU/CNRS (France), MPG (Germany) and IGN (Spain).}}

\author[Fumagalli et al.]{Michele Fumagalli,$^{1}$\thanks{E-mail: mfumagalli@ucolick.org} 
  Miroslava Dessauges-Zavadsky,$^{2}$
  Amy Furniss,$^{3}$\thanks{E-mail: afurniss@ucsc.edu}
  J. Xavier Prochaska,$^{1,4}$\\
  \newauthor
  David A. Williams,$^{3}$
  Kyle Kaplan,$^{5}$ and  Matthew Hogan$^{3}$\\
$^{1}${Department of Astronomy and Astrophysics, University of California,  
1156 High Street, Santa Cruz, CA 95064.}\\
$^{2}${Observatoire de Gen\'eve, Universit\`e de Gen\'eve, 51 Ch. des Maillettes, 1290, Sauverny, Switzerland}\\
$^{3}${Santa Cruz Institute for Particle Physics and Department of Physics, University of California, Santa Cruz, CA 95064, USA}\\
$^{4}${UCO/Lick Observatory, University of California, 1156 High Street, Santa Cruz, CA 95064, USA}\\
$^{5}${The University of Texas, Department of Astronomy, 1 University Station, C1400 Austin, Texas 78712-0259}\\
}

\begin{document}

\date{Accepted xxxx. Received xxxx; in original form xxxx}

\pagerange{\pageref{firstpage}--\pageref{lastpage}} \pubyear{xxxx}

\maketitle

\label{firstpage}


\begin{abstract}
BL Lacertae (Lac) objects that are detected at very-high energies (VHE) are of fundamental 
importance to study multiple astrophysical processes, including the physics of jets, 
the properties of the extragalactic background light and the strength of the 
intergalactic magnetic field. Unfortunately, since most blazars have featureless optical 
spectra that preclude a redshift determination, 
a substantial fraction of these VHE extragalactic sources cannot be used for 
cosmological studies. To assess whether molecular lines are a viable
way to establish distances, we have undertaken a pilot program 
at the IRAM 30m telescope to search for CO lines in three BL Lac objects with known redshifts.
We report a positive detection of $M_{\rm H_2}\sim3\times 10^8$
\msun\ toward \es, but due to the poor quality of the baseline, this value is affected 
by a large systematic uncertainty. For the remaining two sources, \wco\ and \rgb, we
derive $3\sigma$ upper limits at, respectively, $M_{\rm H_2}<8.0\times 10^8$ \msun\
and $M_{\rm H_2}<1.6\times 10^9$ \msun, assuming a line width of 
$150~\rm km~s^{-1}$ and a standard conversion factor $\alpha=4$ \alphaunit.
If these low molecular gas masses are typical for blazars, blind redshift searches 
in molecular lines are currently unfeasible. However, deep observations
are still a promising way to obtain precise redshifts for sources whose 
approximate distances are known via indirect methods. 
Our observations further reveal a deficiency of molecular gas in 
BL Lac objects compared to quasars, suggesting that the host galaxies of these two types 
of active galactic nuclei (AGN) are not drawn from the same parent population.
Future observations are needed to assess whether this discrepancy is statistically significant, 
but our pilot program shows how studies of the interstellar medium in AGN can provide key 
information to explore the connection between the active nuclei and the host galaxies.
\end{abstract}

\begin{keywords}
BL Lacertae objects: general, BL Lacertae objects: individual: 1ES 1959+650,  galaxies: ISM, 
galaxies: distances and redshifts,  galaxies: active, ISM: molecules.
\end{keywords}

\section{Introduction}\label{intro}

Blazars are among the most violently variable radio sources within the population of
active galactic nuclei (AGN).  In the classical unified model \citep[e.g.][]{ant93}, 
they are considered the face-on view of radio-loud AGN, with a beamed jet pointing 
along the observer line of sight and with blobs of highly relativistic particles moving 
down the jet.  The non-thermal radiation 
produces a double-peaked spectral energy distribution (SED), with the lower energy 
peak resulting from the synchrotron radiation of relativistic leptons in the presence of a tangled 
magnetic field, while the higher energy peak is commonly attributed to inverse-Compton 
up-scattering by the relativistic particles within the jet of either the synchrotron photons 
themselves (namely synchrotron self-Compton emission; SSC) or a photon field external to the jet 
(namely external Compton emission; EC). A review of these leptonic models can be found in 
\cite{dermer,maraschi,marscher,sikora} and the references therein.  

Within this general class, 
blazars can be further sub-classified as flat-spectrum radio-loud quasars (FSRQs)
if broad emission lines are visible or, otherwise, as BL Lacertae (Lac) objects.  
BL Lac objects are then categorized based on 
the frequency location of the lower energy peak ($\nu_{\rm synch}$), 
with low-synchrotron-peaked BL Lac (LSP) objects having a $\nu_{\rm synch}$ 
below $10^{14}$\,Hz, intermediate-synchrotron-peaked BL Lac (ISP) 
objects peaking between $10^{14}$ and $10^{15}$\,Hz and high-sychrotron-peaked 
BL Lac (HSP) objects showing a peak above $10^{15}$ Hz \citep{abdoSED}.  
Across these different sub-classes, blazars exhibit a 
continuous variation in their SED \citep{fos98} 
and this has been interpreted as an evolutionary sequence \citep{bot02}
associated to the variation of the diffuse radiation field in the 
surroundings of the relativistic jet \citep{ghi98}. 
The evolution along the sequence is connected to the decrease in the accretion rate and 
to the onset of the radiatively inefficient accretion flow \citep{ho08,tru11}. 
At the beginning of the sequence, blazars have appreciable
external radiation field that facilitates effective cooling, resulting in 
low frequency peaked synchrotron and inverse Compton emission. A
progressively lower contribution of the external field along the sequence inhibits efficient 
cooling, resulting in an higher frequency peak of the synchrotron and inverse Compton emission.

New highly sensitive $\gamma$-ray instruments such as the VERITAS and HESS Cherenkov arrays, 
the \textit{Fermi} Large Area Telescope (LAT), and the MAGIC Telescopes are offering 
a new perspective of the complete electromagnetic spectrum of blazars, providing 
the opportunity to investigate the origin of the highest energy photons
\citep[e.g.][]{ali11}, and to use these powerful AGN
as background sources to probe the properties of 
the extragalactic background light  \citep[EBL; e.g.][]{aha06}.
High energy $\gamma$-ray photons interact with low energy EBL photons via pair 
production, suppressing the intensity of the $\gamma$-ray spectra emitted by 
extragalactic objects. Therefore, the absorbed spectrum contains information on the 
EBL that, in turn, encodes the integrated history of structure formation and the 
evolution of stars and galaxies in the Universe.
Further,  these pair-produced electrons lose energy by inverse Compton 
scattering against the cosmic microwave background radiation. Since the 
resulting flux is dependent on the intergalactic magnetic field (IGMF),
$\gamma$-rays emitting blazars are useful probes of the IGMF  \citep[e.g.][]{ner09}.
However, since the non-thermal emission from BL Lac objects, by definition, outshines 
the weaker optical spectral features that are suitable for establishing redshifts, 
the distances to blazars that are detected at very high energies 
(VHE; $E > 100$ GeV) are sometimes unknown, limiting the use of these objects for 
cosmological studies \citep[e.g.][]{abd11}.

A possibly powerful yet unexplored way to establish the distance to BL Lac objects
is the search for bright carbon monoxide emission lines, ~such as \coone\ and \cotwo. 
Indeed, previous CO observations of quasars and Seyfert I 
galaxies \citep{eva01,sco03,ber07} have yielded a high detection rate 
($\sim 70\%$) of molecular gas from the AGN host galaxies. 
The rate of detection is lower in radio-bright galaxies 
\citep[$\sim 60-40\%$;][]{oca10,smo11}, but one in two or three sources is 
detected in moderately deep observations. Inspired by these previous results, 
we investigate whether molecular emission lines are in fact a promising 
way to measure distances in BL Lac objects. Besides this practical reason, 
establishing the molecular content in BL Lac host galaxies compared to other
classes of AGN potentially offers additional insight into the origin of the 
blazar evolutionary sequence, the unified model 
for AGN and the processes of feedback in massive galaxies. 

This paper presents results from pilot observations of three BL Lac objects 
at the IRAM 30m telescope. The target selection for this 
study is presented in Section \ref{sam}, while observations
and the data reduction are discussed in Section \ref{obs}.
Section \ref{res} offers an overview of the molecular gas properties 
of the three blazars targeted by our observations, while the discussion 
follows in Section \ref{disc}. 
The summary and future prospects conclude this paper in Section \ref{con}.
Throughout this work we adopt cosmological parameters from WMAP7 \citep{kom11}.

\begin{figure*}
  \centering
  \begin{tabular}{ccc}
    \includegraphics[scale=0.25,angle=90]{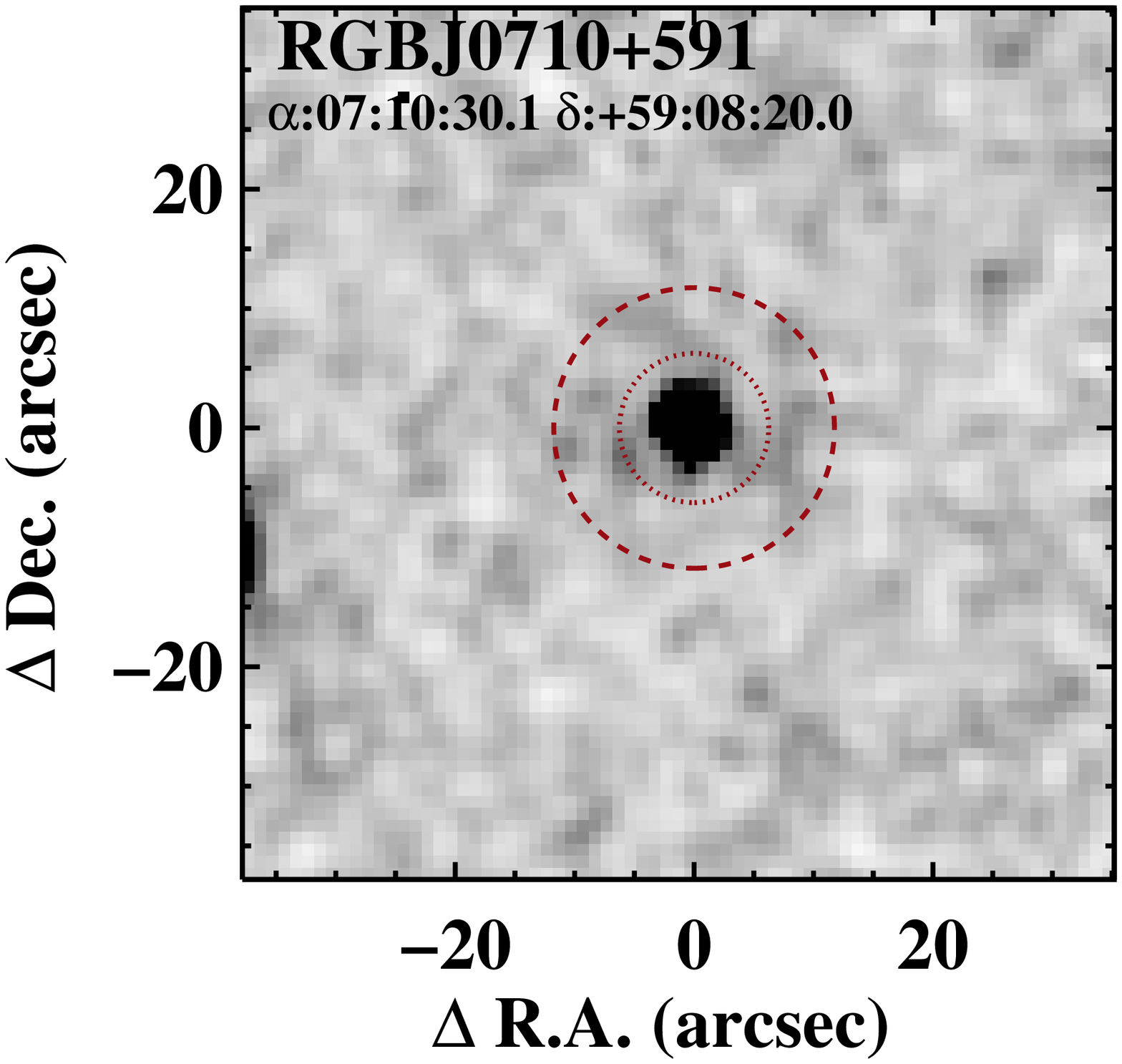}&
    \includegraphics[scale=0.25,angle=90]{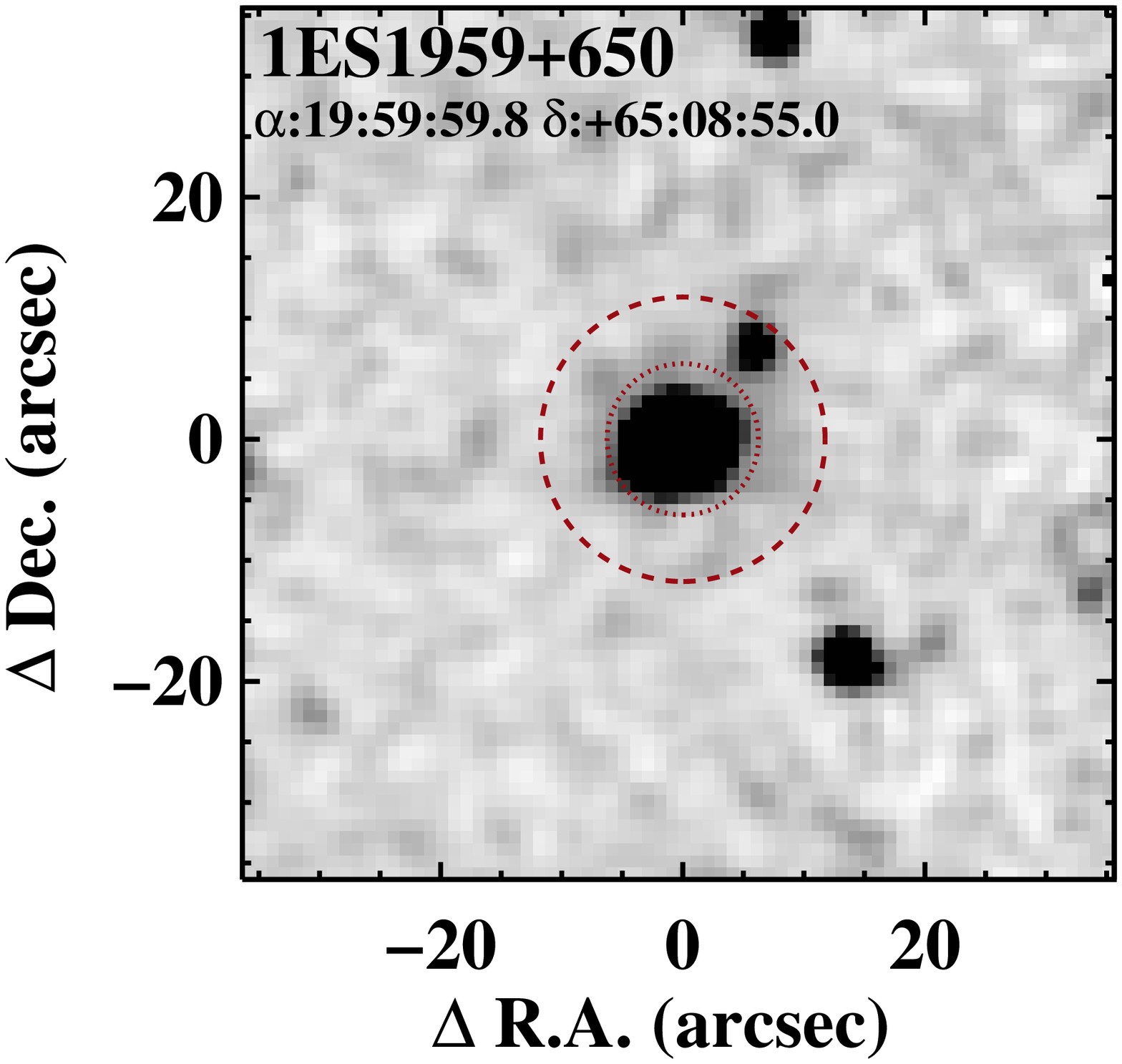}&
    \includegraphics[scale=0.25,angle=90]{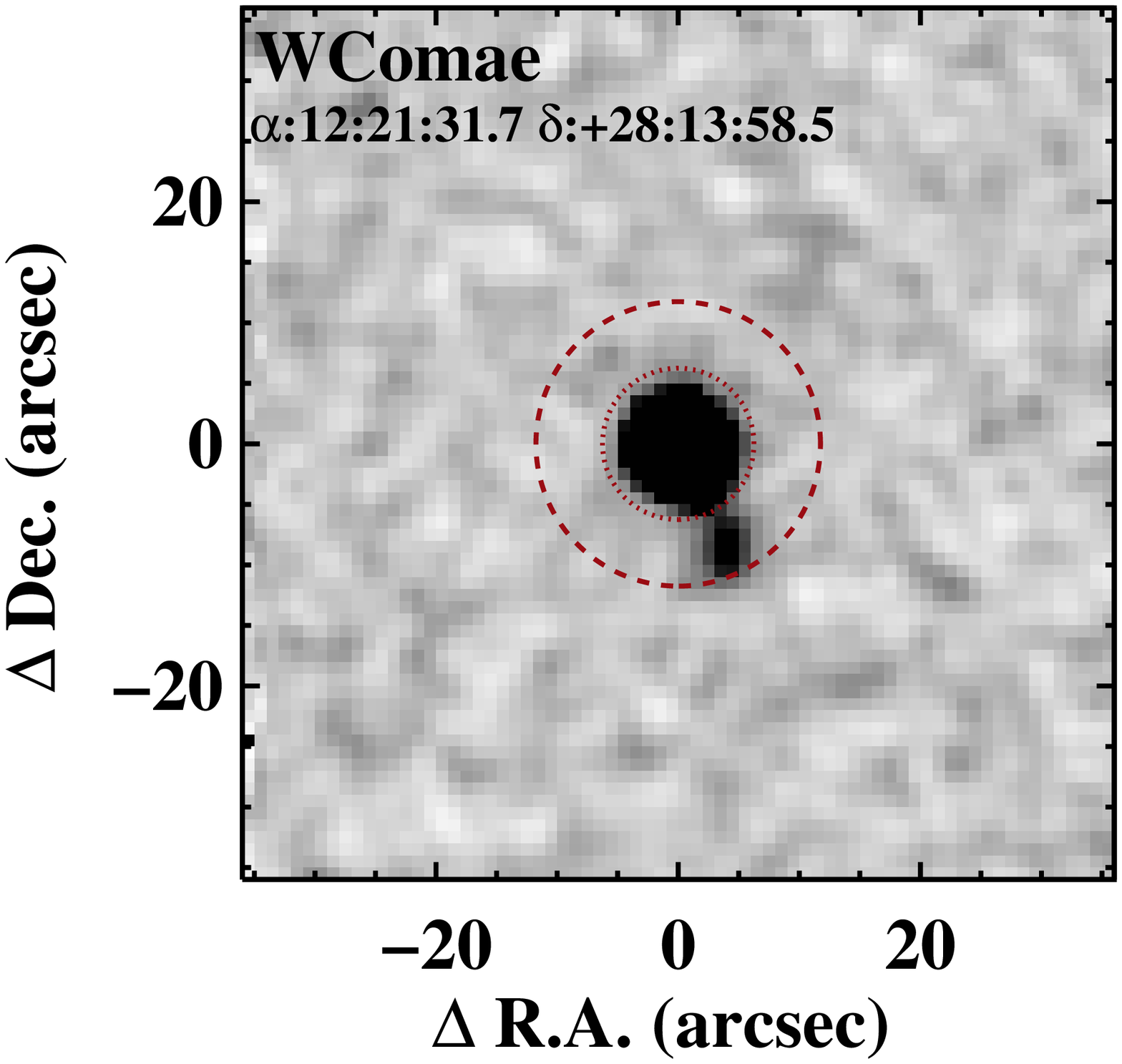}\\
  \end{tabular}
  \caption{$K-$band images
    from the Two Micron All Sky Survey \citep{skr06} for \rgb\ (left), \es\ (center), 
    and \wco\ (right). The two dashed and dotted circles around each blazar 
    represent the beam size at the observed 
    frequencies for the \coone\ and \cotwo\ transitions, respectively. In all cases, 
    the beam encompasses the blazar host galaxies.}
    \label{image}
\end{figure*}

\begin{table*}
\caption{Summary of the molecular emission properties of \rgb, \es, and \wco.}\label{tabco}
\begin{tabular}{l c c c | c c | c c | c c}
\hline
 & & & & \multicolumn{2}{c}{\coone}&\multicolumn{2}{c}{\cotwo}&&\\
Name&$D_{L}$\footnotemark[1]&$z_{\rm opt}$\footnotemark[2]&$z_{\rm co}$\footnotemark[3]&
$T_{\rm rms}$\footnotemark[4]&$I_{\rm co}$\footnotemark[5]&
$T_{\rm rms}$\footnotemark[4]&$I_{\rm co}$\footnotemark[5]&
\lco\footnotemark[6]&\mhmol\footnotemark[7]\\
 &(Mpc)& & & 
 (mK/mJy)&(mJy $\rm km~s^{-1}$)&
 (mK/mJy)&(mJy $\rm km~s^{-1}$)&
 ($10^{7}$ K $\rm km~s^{-1}$ pc$^2$)&($10^{8}M_\odot$)\\
\hline
 \rgb&583&0.125& - &   
 0.20/1.2& $<184$& 
 0.44/3.3& $<491$& 
 $<13.66$ & $<5.47$\\ 
 \es &208&0.047&0.0474&  
 0.18/1.1& $754\pm170^{*}$ & 
 0.44/3.4& $1647\pm511$ & 
 $7.60\pm1.71^{*}$&$3.04\pm0.69^{*}$\\
 \wco&473&0.103& - &   
 0.15/0.9& $<134$ & 
 0.30/2.2& $<334$ & 
 $<6.67$&$<2.67$\\ 
\hline
\end{tabular}
\begin{flushleft}
{\footnotesize $^1$Luminosity distance. $^2$Optical redshift. $^3$Molecular line redshift.
$^4$Root-mean-square antenna temperature and flux density at the final resolution of $\sim 45$ 
    km~s$^{-1}$ per channel. 
$^5$Line intensity, where for non-detections we report the $1\sigma$ limit assuming a 
    line width of $150\rm~km~s^{-1}$. 
$^6$Line luminosity of the \coone\ transition. 
$^7$Molecular gas mass assuming $\alpha=4$\alphaunit. 
$^*$ The line flux and luminosity for the \coone\ transition 
are affected by the poor quality of the baseline (see Section \ref{obs} for details).}
\end{flushleft}
\end{table*}

\section{Target selection}\label{sam}

For this pilot study, we searched the literature for VHE detected
BL Lac objects with a reliable redshift determination and a luminosity distance $D_{\rm L}<600$ Mpc.
Our final selection includes \rgb, \wco, and \es,  three blazars that fall within the
class of radio and X-ray bright sources. Details on the VHE properties of these sources
can be found in \citet{acc09}, \citet{acc10}, and \citet{aha03}.

\rgb\ is an HSP object at $z=0.125$, whose redshift 
is secured by the detection of Ca H \& K and G band absorption lines in the optical spectrum
\citep{lau98}. The blazar host galaxy is fully resolved and the light profile is  
characteristic of an elliptical galaxy \citep{sca00}. 
\es\ is again an HSP object at $z=0.047$ whose redshift is provided by prominent Ca H 
\& K absorption lines \citep{sch93}. The host galaxy is fully resolved and the radial 
profile is very well described by a point source combined with a 
de Vaucouleurs model \citep{sca00}. Finally, \wco\ is an ISP object, whose redshift at 
$z=0.103$ can be established from the emission lines [\ion{O}{II}], [\ion{O}{III}], H$\alpha$, and
[\ion{N}{II}] detected in the Sloan Digital Sky Survey spectrum \citep{aba09}. As 
visible from Figure \ref{image}, a companion galaxy is located in the south-west direction 
from this source, with the two objects possibly in interaction.
The light profile is well described by a S\'ersic index $n\sim 12$ \citep{nil03}.

\begin{figure*}
  \centering
  \begin{tabular}{cc}
    \includegraphics[scale=0.33,angle=90]{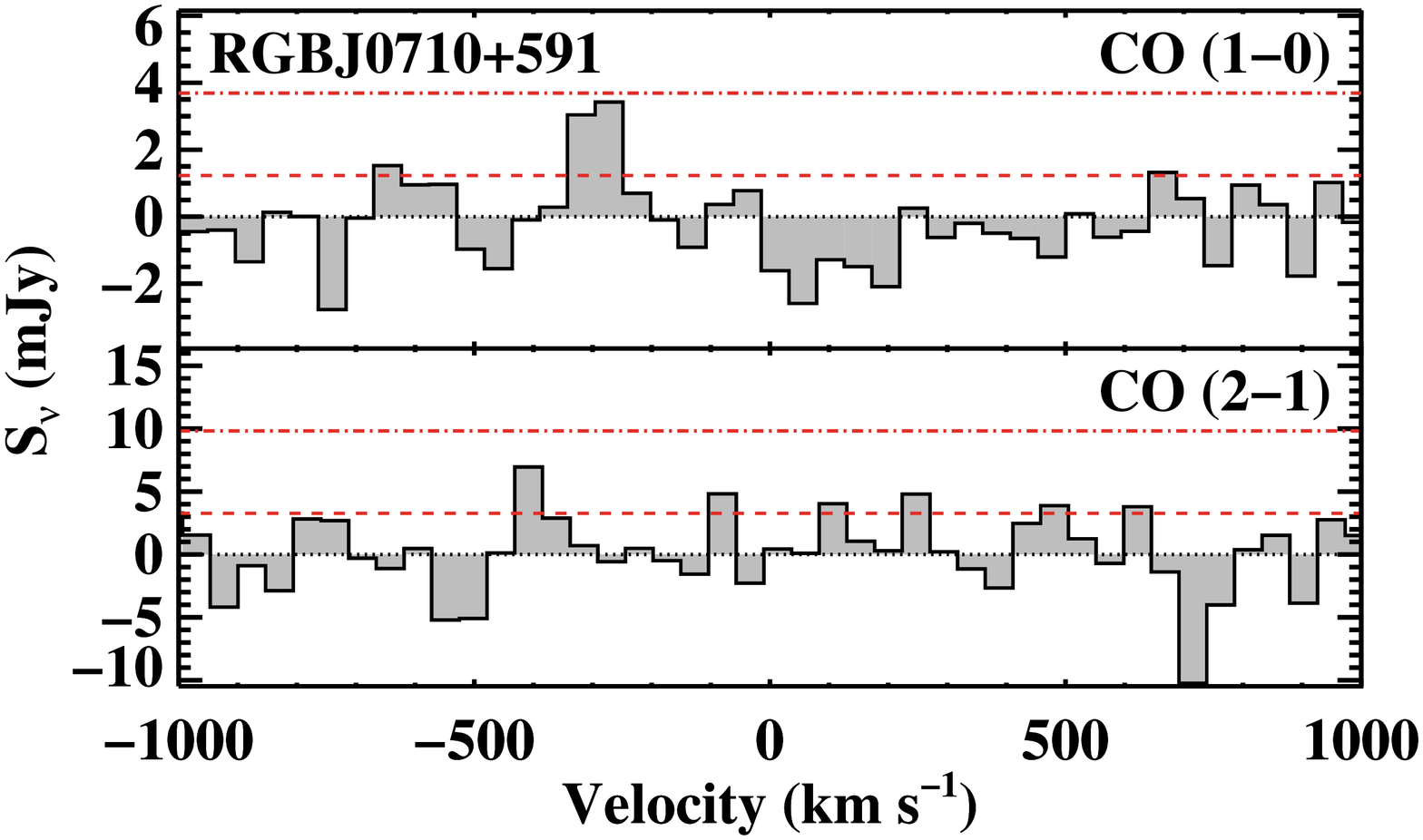}&
    \includegraphics[scale=0.33,angle=90]{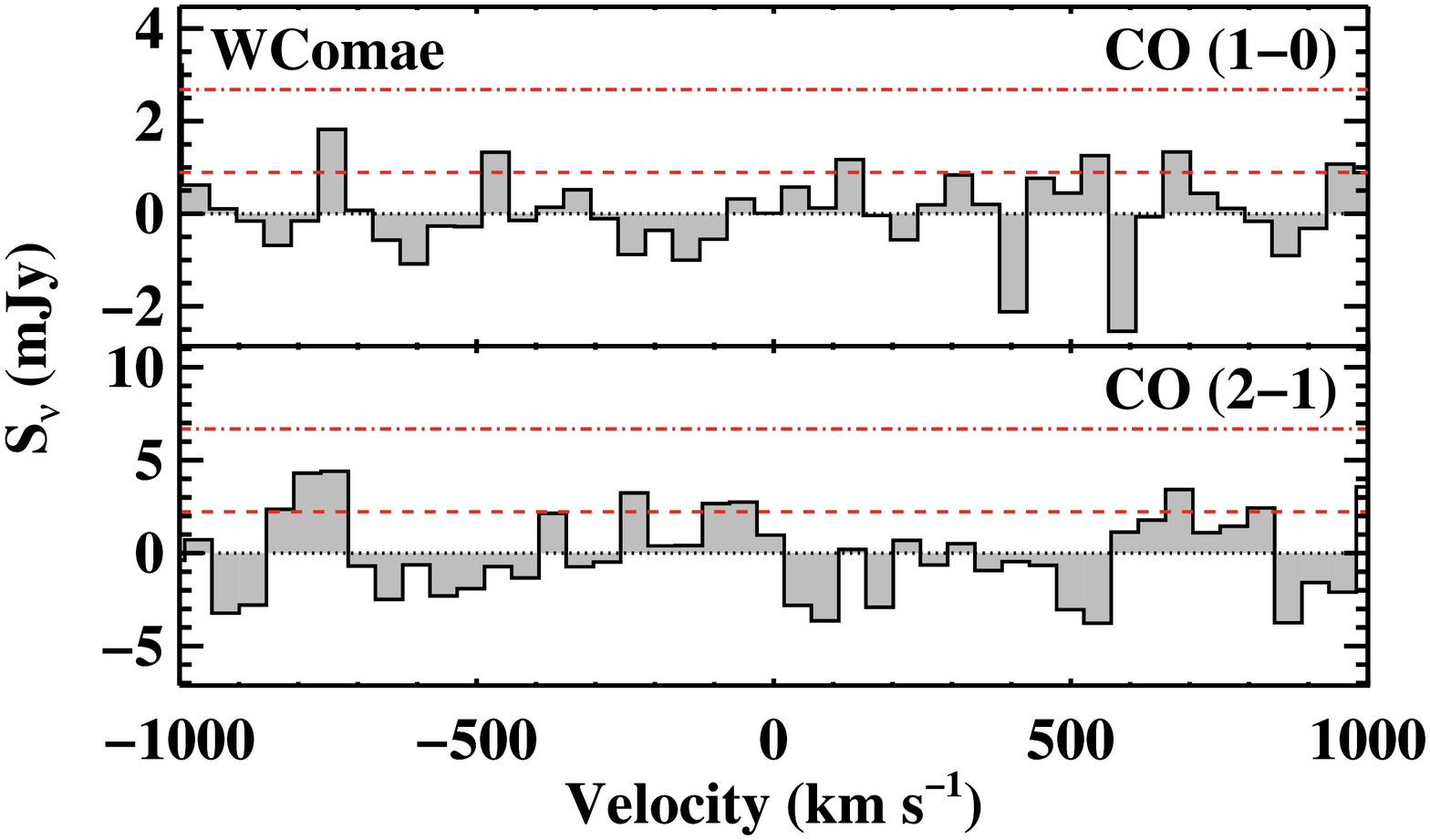}\\
    \includegraphics[scale=0.33,angle=90]{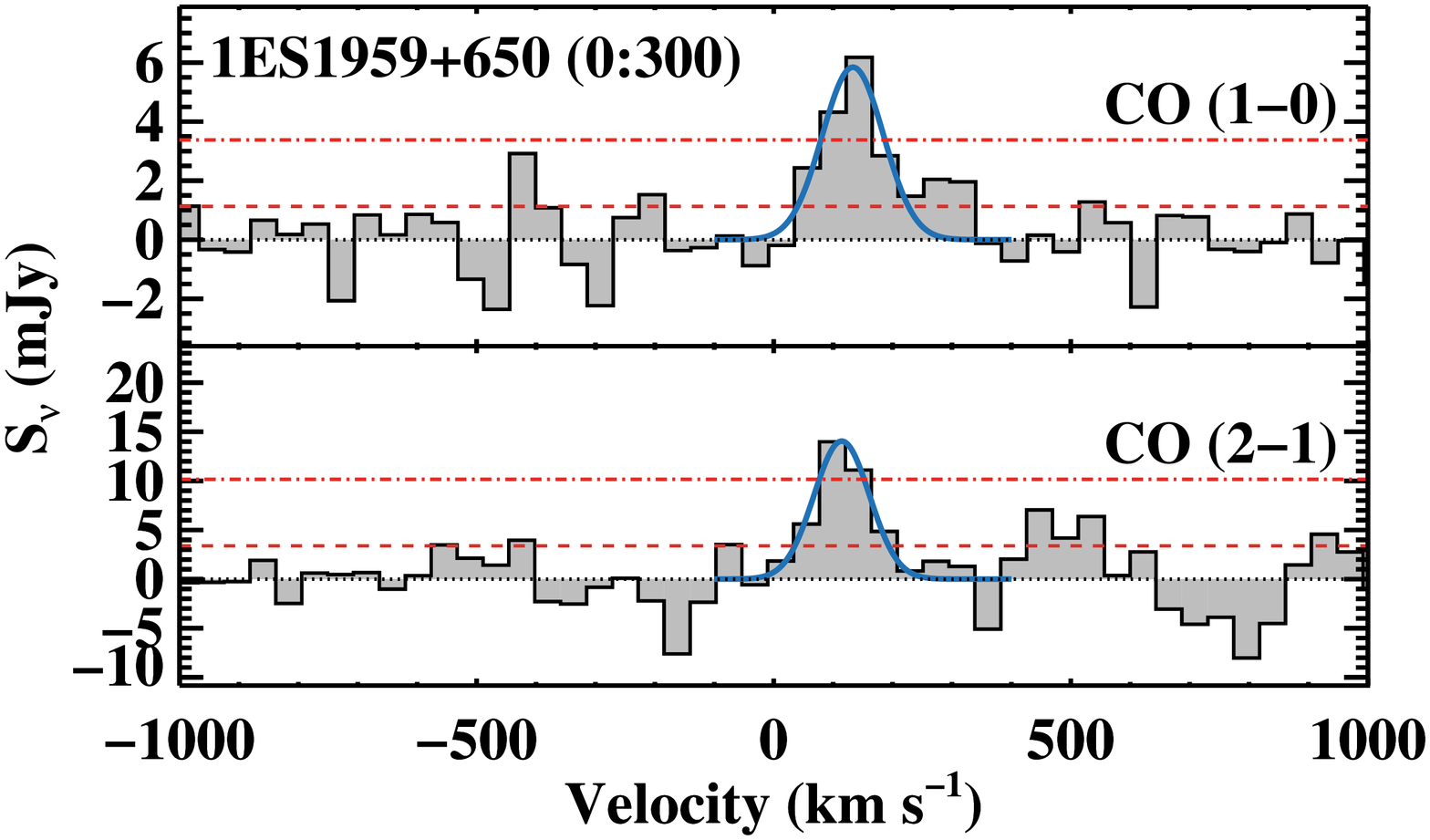}&
    \includegraphics[scale=0.33,angle=90]{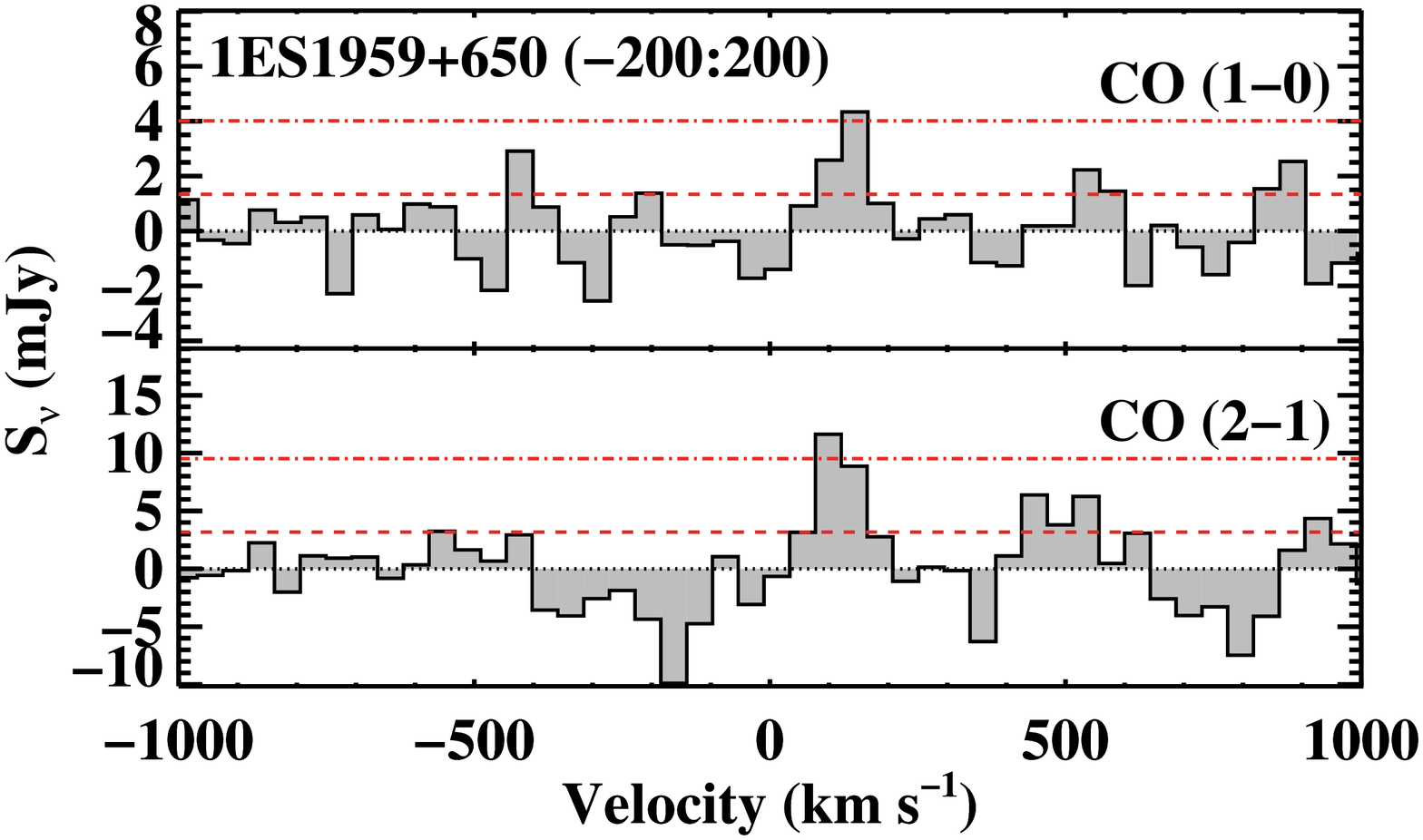}\\
  \end{tabular}
  \caption{Spectra of the \coone\ and \cotwo\ transitions, centered at the systemic velocity
    that is known from optical spectroscopy. The dashed and dash-dotted horizontal 
    lines mark one and three times the rms of each spectrum. For \es\ (bottom panels), 
    we show two different spectra that have been obtained with a fine-tuned window
    in the velocity interval $0-300~\rm km~s^{-1}$ (left) and a symmetric window 
    $\pm 200~\rm km~s^{-1}$ (right). Gaussian models for the detected lines in \es\ are 
    also shown with blue solid lines.}
  \label{spec}
\end{figure*}

\section{Observations and data reduction}\label{obs}

We observed the three blazars with the IRAM 30m telescope on Pico Veleta (Spain)
between UT 15 and 19 July 2011. Additional observations for \rgb\ were taken 
between UT 2 and 5 September 2011. Weather conditions were generally good, 
with system temperatures ranging between 100 and 150 K at $\sim 110$ GHz.

To acquire the \coone\ line (rest-frame frequency 
$\nu_{\rm 0}=115.2712018$ GHz) and the \cotwo\ line ($\nu_{\rm 0}=230.5380000$ GHz) simultaneously, 
we used four single pixel heterodyne EMIR receivers, two centered on 
the E0 band ($\rm 3~mm$) and two on the E2 band ($\rm 1~mm$). The local oscillator was 
tuned such that the redshifted \coone\ lines would fall at the center of the lower inner 
sideband for \rgb\ and \wco\ and, due to its lower redshift, in the upper inner sideband for \es. 
The telescope half-power beam widths are $\sim 23''$ and $\sim 12''$, respectively at 
the \coone\ and \cotwo\ redshifted frequencies, and encompass the host galaxies 
(Figure \ref{image}). The data were recorded using the WILMA 
autocorrelator providing a spectral resolution of $\rm 2~MHz$. This corresponds to a resolution of 
$\sim 6~\rm km~s^{-1}$ at $\rm 3~mm$ and $\sim 3~\rm km~s^{-1}$ at $\rm 1~mm$. The observations were
conducted in wobbler-switching mode, with a switching frequency of $\rm 0.5~Hz$ and a symmetrical 
azimuthal wobbler throw of $50''$ to maximize the baseline stability. 
For each blazar, we acquired series of 12 ON/OFF subscans of 30 seconds each. 
Calibrations were repeated every 6 minutes, while focus and pointing were checked 
throughout the observing period with bright sources. At the end of the observing run, 
we had acquired 4.8 hours on source for \es, 7.6 hours for \wco\ and 5.4 hours for \rgb.

The data reduction was completed with the IRAM software CLASS/GILDAS. The flux calibrated 
scans were subsequently combined and analyzed using a set of in-house IDL procedures. 
To compensate for the variability of the continuum level that is common for blazars, 
we subtracted a constant baseline from each scan before coadding. After applying the heliocentric 
correction to homogenize observations at different epochs, we combined all the scans, 
weighting by the exposure time and the system temperature. Both polarizations were included 
in the final stack. 

Due to the bright continuum level at $\rm 3~mm$ for \es\ and \wco\ ($\sim 0.16$ Jy 
and $\sim 0.23$ Jy, respectively), the quality of the baseline is degraded 
by the presence of standing waves between the secondary and the feed to the receiver.
Specifically, the spectrum of \es\ at  $\rm 3~mm$ exhibits ripples with a peak-to-peak 
amplitude of $\sim 2~\rm mK$ over characteristic velocities of $\sim 500~\rm km~s^{-1}$.
Similarly, the spectrum of \wco\ shows flux variations of $\sim 0.5~\rm mK$ over
characteristic velocities of $\sim 500~\rm km~s^{-1}$. 
For this reason, we did not attempt to model the baseline over the entire 4 GHz bandwidth, 
instead we focused our analysis on a velocity window of $\pm 1000~\rm km~s^{-1}$ centered 
at the systemic redshift that is known from optical spectroscopy. 
In this interval, we subtracted a polynomial model fitted to the data from the spectrum. 
Then, we smoothed each spectrum with a Hann window and we binned the data to achieve a
final resolution of $\sim 45~\rm km~s^{-1}$ per channel. Figure \ref{spec} shows the smoothed 
continuum-subtracted spectra for the three blazars. 

For \rgb, we subtracted a linear baseline for both the \coone\ and the \cotwo\ transitions.
While fitting the baseline, we excluded a symmetric window of $\pm 200~\rm km~s^{-1}$ 
centered at the systemic redshift to prevent the subtraction of weak emission lines. 
A higher order polynomial fit (order 10 for the \coone\ line and order 5 for the \cotwo\ line)
was required instead for \wco\ due to the lower quality of the baseline,
again excluding a symmetric window of $\pm 200~\rm km~s^{-1}$ around the optical redshift. 
We emphasize that, despite the high order adopted, the model for the baseline has 
characteristic shapes on scales of several hundred $\rm km~s^{-1}$ and, therefore, 
it does not introduce spurious signals on small velocity scales 
($<200-300$ km~s$^{-1}$), preserving the noise properties of the original spectrum.  
Finally, for \es, the \cotwo\ emission line is visible at $\sim 100~\rm km~s^{-1}$ 
even before modeling the baseline. In this case, we adopted a polynomial fit of 
order 4 with a $0-300~\rm km~s^{-1}$ window. Conversely, the poor quality of 
the baseline at longer wavelengths  prevents us from detecting the associated 
\coone\ line trivially. After subtracting 
a baseline of order 15, a line is clearly detected when,
based on the position of the \cotwo\ line, a $0-300~\rm km~s^{-1}$ window
is excluded from the continuum fit (bottom left panel of Figure \ref{spec}). 
Instead, if we impose a symmetric window of $\pm 200~\rm km~s^{-1}$ around the optical redshift
as previously done for the other two blazars, the \coone\ line 
is only marginally detected and the peak flux is suppressed by $\sim 65\%$ 
(bottom right panel of Figure \ref{spec}). For this 
reason, we consider the \coone\ line flux affected by significant systematic 
uncertainty associated with the baseline subtraction.
 
The root-mean-squared temperatures ($T_{\rm rms}$) computed in the baseline-subtracted 
spectra over the $\pm 1000~\rm km~s^{-1}$ velocity interval at the final resolution are 
listed in Table \ref{tabco}, in units of both 
the antenna temperature $T^*_A$ and of the flux density $S_\nu$. For the flux calibration, 
we adopt the frequency dependent conversion $S/T^*_A$ available from the EMIR commissioning report.

\begin{figure*}
  \centering
  \includegraphics[scale=0.45]{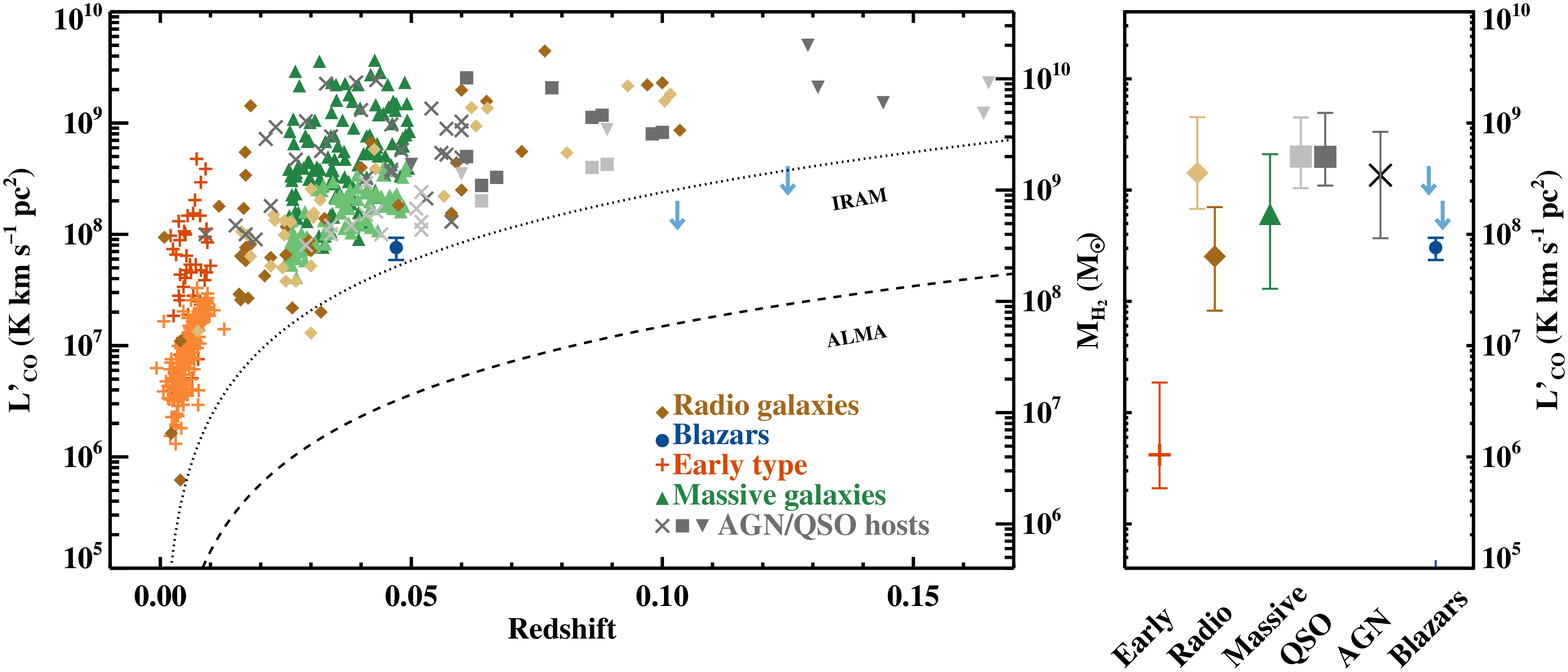}
  \caption{\emph{(Left)} \coone\ line luminosity as a function of redshift in: 
    early type galaxies \citep[orange crosses;][]{you11},
    massive galaxies \citep[green triangles;][]{san11}, quasars \citep[gray squares;][]{sco03},
    low luminosity quasars \citep[gray X's;][]{ber07}, infrared-excess quasars 
    \citep[downward triangles;][]{eva01}, and radio galaxies \citep[diamonds;][]{oca10,smo11}. 
    Dark colors are for detections, while lighter 
    colors are for the $3\sigma$ upper limits. The blue circle marks the line luminosity
    for \es, while blue arrows indicate the upper limits for \wco\ and \rgb. 
    The right-end-side axis translates the \coone\ luminosity into the molecular hydrogen mass, 
    assuming $\alpha=4$ \alphaunit. The  black dotted and dashed lines represent the luminosity 
    at constant flux density as a function of redshift for the IRAM 30m telescope 
    and for ALMA. \emph{(Right)} The 25$^{\rm th}$, 50$^{\rm th}$, and 75$^{\rm th}$
    percentiles for individual classes, computed including upper limits.
    The percentiles computed for a subset of the
    quasars and radio-loud galaxies are also shown with lighter colors 
    to assess the importance of selection effects (see main text). 
    BL Lac host galaxies appear to have lower molecular gas than quasars.}
  \label{lim}
\end{figure*}

\section{The molecular content of BL Lac objects}\label{res}

Figure \ref{spec} shows that molecular gas is only detected in \es, i.e. in one out of three 
sources. Inspecting more closely the \coone\ spectrum for \rgb, a positive excess 
is visible at $-300\rm~km~s^{-1}$. Integrated over two channels, this spectral feature 
is significant at $< 2\sigma$ after accounting for both the statistical error
and the uncertainty of the baseline, which we estimate to be comparable to the 
measured $T_{\rm rms}$. For this reason, and because we do not see the 
corresponding \cotwo\ transition, we only report an upper limit for \rgb.

More significant, instead, is the detection for \es. The \cotwo\ line is detected 
at $3.2\sigma$ accounting for both random and systematic errors, 
and it is not particularly sensitive 
to the baseline subtraction. Further, the \coone\ line is also detected at $4.4\sigma$, but as 
previously discussed this transition is more sensitive to the baseline subtraction.
Moreover, both lines appear at the same redshift ($z_{\rm co}=0.0475\pm0.0002$ 
and $z_{\rm co}=0.0474\pm0.0002$ for, respectively, \coone\ and \cotwo) that 
is consistent with the one measured from optical lines. Finally, both line 
profiles are marginally resolved.

For \es, we model both lines with a Gaussian profile in the velocity 
interval $0-250~\rm km~s^{-1}$. The peak of the molecular
emission is offset from the systemic velocity inferred from optical 
spectroscopy by $\sim 130~\rm km~s^{-1}$, 
but since the uncertainty of the optical redshift is unknown, this shift 
may not be significant. The  \coone\ and \cotwo\ lines have a full-width half-maximum 
velocity  $\Delta v=121~\rm km~s^{-1}$ and $\Delta v=110~\rm km~s^{-1}$, respectively.  
These lines are narrow compared to the typical line widths for luminous quasars \citep{sco03}
and more closely resemble the typical velocity width of early-type galaxies \citep{you11}.
By integrating over the Gaussian profile, we derive a line intensity\footnote{By directly 
integrating the data in the velocity interval $0-250~\rm km~s^{-1}$, we find similar 
values, consistent within the errors.}
$I_{\rm co}=754\pm170$ mJy~km~s$^{-1}$ for \coone\ and 
$I_{\rm co}=1647\pm511$ mJy~km~s$^{-1}$ for \cotwo. Here, the error is computed as
$\delta I_{\rm co} \propto \sqrt{2N} T_{\rm rms} v_{\rm res}$, where $N$ is the number of channels 
at resolution $v_{\rm res}$ within the line width.
We also attempt to estimate the uncertainty on the baseline by performing 
multiple fits to the continuum with varying choices of polynomial orders and 
window positions. This simple exercise reveals that an additional factor 
$\sqrt2$ should be included in the error budget. However, we emphasize that 
this error estimate is still sensitive to the functional form adopted to model 
the ripples that are present in the baseline, and thus an additional systematic 
uncertainty may affect our final values. Only future interferometric 
observations will be able to quantify this error.

Finally, the integrated  intensity for the \coone\ line 
can be converted into the luminosity \lco\ (in \lcounit) by using
\begin{equation}
L'_{\rm CO}=2.45\times 10^3 \frac{I_{\rm co}}{\rm Jy~km~s^{-1}}
\Big(\frac{D_{\rm L}}{\rm Mpc}\Big)^2 \frac{1}{1+z}\:.
\end{equation}
Assuming a typical conversion factor from CO luminosity to molecular hydrogen mass 
$\alpha=4$\alphaunit\ \citep[see][]{eva01}, we infer a total hydrogen mass 
$M_{\rm H_2}=(3.04\pm0.69)\times10^8$ M$_\odot$.
A summary of the molecular gas properties of \es, together with the $1\sigma$ upper limits
for \rgb\ and \wco, is provided in Table \ref{tabco}.

\section{Discussion}\label{disc}

With two tight upper limits and one detection 
at $M_{\rm H_2} \sim 3 \times 10^{8} \rm~M_\odot$,
these pilot observations suggest that VHE-detected BL Lac objects do not generally bear 
large amounts of molecular gas. However, the detection of weak lines in \es\ proves that, 
despite the bright synchrotron continuum, deep enough observations of the molecular 
lines can be a viable technique to constrain and
refine the redshifts of sources of particular interest for which approximate distances 
are know from indirect methods (see Section \ref{con}). 

\subsection{A comparison with other AGN and galaxy populations}

Although our sample is too small for any conclusive analysis, 
in the left panel of Figure \ref{lim} we attempt a first comparison between 
the CO luminosity of BL Lac objects and other sources for which millimeter 
observations are available from the literature. All the data have been homogenized to a 
single conversion factor $\alpha=4$\alphaunit\ and all the limits are reported at 
$3\sigma$, preserving the choice of the velocity width
as published in the original studies. For the upper limits of \rgb\ and \wco, we adopt a 
line width of $150\rm~km~s^{-1}$ that is motivated by the detection in \es. 

The difference between VHE-detected BL Lac objects and quasars \citep{sco03,eva01} or Seyfert I 
host galaxies \citep{ber07} is evident, given that the molecular gas detected in 
quasars is up to a factor of ten higher than what was found in blazars by our observations.
This discrepancy is apparent even if we consider only quasars from 
the \citet{sco03} sample that  overlap in redshift with our blazars and that 
were selected purely based on optical luminosity. In fact, there is an even more pronounced 
separation between blazars and infrared ultra-luminous quasars that have a 
median molecular gas mass of $6\times 10^9$ \msun\ \citep{xia12}.

Further, BL Lac objects are only consistent with the lower-end of the molecular gas 
distribution in galaxies that are selected purely based on their stellar mass 
($M_{\rm star}>10^{10}$\msun) regardless to the presence of AGN and morphological type \citep{san11}.
Conversely, the CO luminosity in BL Lac objects seems consistent with the level of 
molecular gas typically detected in radio-loud galaxies. These sources are indeed  
observed to have a lower molecular gas content than quasars \citep{oca10,smo11}.
Finally, VHE-detected BL Lac objects also appear compatible with the population of 
early type galaxies \citep{you11}
in which only a fifth of the observed sources are detected at the current sensitivity limits. 
It should be noted that these classes of objects are observed at lower redshifts
than the blazars and the luminous quasars, thus the  Malmquist bias complicates a
systematic comparison between different classes. For instance, the subset of radio 
galaxies with $z\gtrsim 0.05$ exhibits a molecular mass that is 
systematically higher and comparable to quasars than in the 
full sample \citep[see additional discussion on this redshift dependence in][]{oca10,smo11}.
Nevertheless, the comparison between quasars and BL Lac objects is done at comparable
redshifts. Further, because blazars lie at $z\gtrsim 0.05$, we can still
conclude that blazars are consistent with the median molecular mass content of early type 
galaxies and radio galaxies given our current sensitivity limit. Deeper observations
will investigate whether this is true or whether blazars exhibit in fact an intermediate
level of molecular gas between those of quasars and early type galaxies.

The qualitative trends discussed above are confirmed by the more 
quantitative analysis presented in 
the right panel of Figure \ref{lim}, where for each class 
we compare the 25$^{\rm th}$, 50$^{\rm th}$, and 75$^{\rm th}$ percentiles of the 
observed \lco\ distributions. For this, we use the Kaplan-Meier estimator \citep{fei85} 
for randomly censored  data as implemented in the ASURV package \citep{lav92}. Since our 
sample includes only three blazars, percentiles cannot be defined and we superimpose 
the observed data points in Figure \ref{lim}.
We can still derive a mean value of \lco$=7.6\times 10^{7}$ \lcounit\ that,
however, is not particularly informative due to small statistics.
We further emphasize that the percentiles shown in the right panel of Figure \ref{lim}
refer to samples that are selected to be representative of the populations in exam,
although they are not complete.

\subsection{Possible observational biases}

Noting the lack of strong emission from these BL Lac objects, it is worth examining whether the
discrepancy between this class of blazars and quasars can be attributed to 
an observational bias and, in particular, whether the bright synchrotron continuum may 
outshine strong emission lines. 

To rule out this possibility, we constructed mock spectra with different continuum intensities
and line properties similar to the one measured in the \coone\ transition for \es. 
Following the same procedure adopted for real data, we analyzed these mock spectra multiple 
times, adding different realizations of Gaussian noise to a level that is comparable 
to the observed one.
At the end of this test, we were able to recover the input line parameters with a dispersion that is 
consistent with the estimated uncertainties. 
We also built an additional set of mock spectra, with the noise properties 
inferred for \rgb. Superimposing to 
the continuum a line of total intensity equal to three times  $T_{\rm rms}$, we were able to visually 
identify the emission line. We conclude that, if present, emission lines significantly 
brighter than the quoted limits should have been detected despite the bright synchrotron continuum. 
This procedure does not take into account the ripples that are present
at $\rm 3~mm$ in the spectra of \wco\ and \es, and future interferometric observations will 
assess the error introduced by the poor quality of the baseline.

Next, we consider whether the different luminosities in BL Lac objects and quasars can be 
attributed to projection effects. Since the emitted flux from optically thick 
regions is proportional to the gas surface area, different viewing angles 
of the same source can result in different \lco. Assuming that the gas settles in a
disk  and that the jets extend above the plane, observations 
of blazars probe the face-on view of the emitting disk. Thus, in order to explain the 
lower luminosity of BL Lac objects, the gas should be concentrated in a small region
where the disk surface area is comparable or smaller than the disk scale height. However, 
this seems implausible for at least two reasons. First, a concentration of $10^8$ \msun\ 
of molecular gas in the inner few hundred parsecs close to the black hole 
would produce a hydrogen density that exceeds what is inferred from the narrow and broad line 
regions \citep{ost06}, by at least two orders of magnitude. Second, the few 
observations of quasars where the gas disk is resolved show that the 
emitting regions typically extend to $\sim 1$ kpc or, in some cases, 
up to a few kpc \citep{rie11}. 

\subsection{On the difference between quasars and BL Lac objects}

Considering morphological studies of blazars \citep[e.g.][]{sca00,urr00},
a lack of molecular gas is not completely surprising since 
almost all the BL Lac objects resides in massive early-type galaxies. 
These studies further suggest that the nuclear activity is not
necessarily triggered by recent interactions that have a 
dramatic effect on the large-scale morphological and structural properties of the host galaxies.
In fact, radio galaxies, blazars and radio-quiet elliptical galaxies share similar 
structural and morphological characteristics. Our CO observations add  additional 
evidence in support to this idea: in our limited sample the cold gas in BL Lac objects 
is consistent with the lower level of CO seen in early-type and radio galaxies
(see Figure~\ref{lim}).

Further,  low-excitation radio AGN 
(LERAGN) and Fanaroff and Riley class I (FR-I) sources 
have a lower  molecular gas content than high-excitation 
radio AGN (HERAGN) and FR-II sources \citep{smo11,oca10}. 
Although it is still debated whether the discrepancy between FR-I and FR-II sources is due to 
selection effects \citep{oca10}, the low molecular content found in BL Lac objects 
is consistent with these findings, if we allow for an admittedly loose 
identification of LERAGN with FR-I class, which contains most of the BL Lac population. 
Thus, based on the properties of their interstellar medium, low-redshift quasar and
BL Lac host galaxies appear as two distinct families \citep[see also][]{sco03}
that are not drawn from the same parent population.

In light of recent findings that both the amount of atomic interstellar medium  \citep[ISM;][]{fab11}
and the bulk velocity of the molecular content in nearby AGN hosts \citep{gui12} are only weakly 
affected by the nuclear activity, the low molecular content 
in BL Lac objects might be a consequence of the quenching mechanisms associated with the
growth of bulges and an increase in the stellar surface density \citep{kau12}
rather than a direct consequence of the presence of a jet.  In any case, 
regardless of whether the AGN are causally linked to the gas properties of the host 
galaxies \citep[e.g.][]{nes11}, the ISM content of BL Lac objects revealed by our observations 
is typical of late stages of galaxy evolution, as it is also advocated to
explain the blazar sequence \citep[e.g.][]{ghi98,bot02}: 
BL Lac objects correspond to late stages of activity in which the central 
black hole is fed by a radiatively inefficient accretion disk. 
This view is further supported by the fact that VHE emitting HSP BL Lacs are generally
found to be best represented by simple SSC models, while VHE emitting 
ISP and LSP objects require external photon fields to describe their broadband data.  
Moreover, mid-IR observations of a 
large population of BL Lac objects show no clear signatures of torus emission,
consistent with the expected low accretion rates \citep{plo12}. Thus, our observations fit within 
unified models for AGN in which the accretion rate shapes the observed nuclear
properties. BL Lac objects are not simply AGN with radio emission superimposed, 
as described in unification models that are based solely on geometrical obscuration \citep{ant93},
but are objects in which the onset of the radiatively inefficient accretion flow
has led to the disappearance of the broad emission lines \citep[e.g.][]{ho08,tru11}.

In closing, we note that the discrepancy in CO luminosity between 
BL Lac and quasar host galaxies suggested by our observations has consequences
that extend beyond the study of AGN. It has recently  been noted that 
plasma instabilities that develop in the jets of TeV-emitting blazars could inject 
a considerable amount of heat in the intergalactic medium \citep{bro11}, altering the thermal 
history of the Universe at low redshifts and at low densities \citep{cha11}.
However, if BL Lac objects and quasars are not drawn from the same parent population, 
the number density of blazars and quasars may not evolve in parallel 
across all redshifts.  Thus, the underlying assumption that the TeV blazar 
activity does not lag that of quasars may need additional verification. 
Additionally, observations of 
high-excitation radio galaxies show a stronger cosmic evolution at 
all luminosities compared to the low-excitation radio galaxies which exhibit milder 
evolution \citep{wil01,sad07,don09,smo09,bes12}. A similar trend is also expected 
for the relative evolution of FSRQs and BL Lac objects \citep{cav02,bot02}.
Besides this specific application, understanding the connection between the 
heating mechanisms of blazars and the ISM properties of the low-excitation radio 
galaxies becomes relevant in the context of current feedback prescriptions such 
as the ``radio-mode'' feedback that is commonly adopted in 
semi-analytic models \citep[e.g.][]{cro06,bow06}.

\section{Summary and Future prospects}\label{con}

BL Lac blazars detected at VHE are key sources to probe a variety of 
astrophysical processes, including the physics of jets, the properties of the
EBL and the strength of the IGMF.
However, due to their featureless optical spectra, it is often impossible to establish 
a redshift with 
optical emission or absorption lines, limiting the use of these objects in cosmological studies. 
Motivated by the need to reliably determine distances for VHE BL Lac objects
with other techniques, we have undertaken a pilot program at the IRAM 30m telescope 
to assess the molecular gas content in three sources with known redshifts
to establish whether molecular lines can be used to trace distances for this class of blazars. 

For one BL Lac object (\es), we detected both the \coone\ and the \cotwo\ transitions 
at $>3\sigma$ confidence level, corresponding to a molecular mass of $M_{\rm H_2}\sim3\times 10^8$
\msun. However, due to the poor quality of the baseline, we regard this detection as 
affected by a large systematic uncertainty. Conversely, for the other two sources 
(\wco\ and \rgb), we do not detect CO emission to $3\sigma$ limits \lco$<2\times 10^8$ 
\lcounit\ and \lco$<4\times 10^8$ \lcounit, respectively. 
In both cases, we assumed a line width of  $150~\rm km~s^{-1}$, motivated by the 
lines detected in \es. Assuming a conversion factor $\alpha=4$\alphaunit, 
these limits translate to $M_{\rm H_2}<8\times 10^8$\msun and $M_{\rm H_2}<1.6\times 10^9$\msun.

Although this sample is too small to derive robust statistical conclusions, our observations 
suggest that BL Lac host galaxies may not, in general, 
bear large amounts of molecular gas. The detected 
levels are in fact comparable to what is observed in 
early-type or radio galaxies. Thus, a search for molecular lines is not a feasible 
approach to establish blazar redshifts with instruments that offer a limited bandwidth.   
Nevertheless, current facilities such as the IRAM Plateau de Bure Interferometer or the 
Atacama Large Millimeter Array can reach up to one order of magnitude better sensitivity
than what is achievable in similar integration times with the IRAM 30m telescope (see Figure \ref{lim}).
Therefore, the detection of molecular lines can still be a promising way to obtain precise 
redshifts for blazars whose approximate distances have already been  estimated through 
indirect methods such as the absorption from the intervening Ly$\alpha$ forest \citep[e.g.][]{dan10},
the EBL attenuation \citep[e.g.][]{ste10}, or the photometric
properties of host galaxies \citep[e.g.][]{sba05}.     

Besides the possibility of establishing redshifts, these observations offer
tantalizing hints that the molecular gas content of BL Lac objects is 
lower than that of quasars. Larger samples are now needed to assess the significance 
of this discrepancy, but a systematic deficiency of molecular gas in BL Lac objects 
compared to quasars will provide confirmation that the relative host galaxies are 
not drawn from the same parent population.
Our observations, together with previous studies of blazars and radio sources 
(e.g. the low and high excitation radio AGN or the FR-I and FR-II sources), 
provide increasing evidence that BL Lac objects represent later stages of galaxy evolution, 
in which the central black hole is fed by a radiatively inefficient accretion disk.
In the era of large millimeter arrays, a multiwavelength view of the ISM of 
AGN is required to fully explore the connection between different types of active nuclei and 
the host galaxies.

\section*{Acknowledgments}
We would like to thank the IRAM staff for the tremendous help provided 
during the observations. We are particularly grateful to F. Walter, 
M. Dotti, J. Trump, and A. Treves for comments on this manuscript. We acknowledge 
useful discussions with E. Ramirez, M. MacLeod, and M. Colpi.
This work was supported in part by the US National Science Foundation grant PHY09-70134.

\label{lastpage}

\begin{thebibliography}{}
\bibitem[Abazajian et al.(2009)]{aba09} Abazajian, K.~N., 
  Adelman-McCarthy, J.~K., Ag{\"u}eros, M.~A., et al.\ 2009, \apjs, 182, 543 
\bibitem[Abdo et al.(2011)]{abd11} Abdo, A.~A., Ackermann, 
  M., Ajello, M., et al.\ 2011, \apj, 726, 43 
  \bibitem[Abdo et al.(2010b)]{abdoSED} Abdo, A. et al.\ 2010, \apj, 716, 30
\bibitem[Acciari et al.(2010)]{acc10} Acciari, V.~A., Aliu,    
  E., Arlen, T., et al.\ 2010, \apjl, 715, L49 
\bibitem[Acciari et al.(2009)]{acc09} Acciari, V.~A., Aliu,   
  E., Aune, T., et al.\ 2009, \apj, 707, 612 
\bibitem[Aharonian et al.(2006)]{aha06} Aharonian, F., 
  Akhperjanian, A.~G., Bazer-Bachi, A.~R., et al.\ 2006, \nat, 440, 1018 
\bibitem[Aharonian et al.(2003)]{aha03} Aharonian, F., Akhperjanian, A.,    
  Beilicke, M., et al.\ 2003, \aap, 406, L9 
\bibitem[Aliu et al.(2011)]{ali11} Aliu, E., Aune, T., 
  Beilicke, M., et al.\ 2011, \apj, 742, 127 
\bibitem[Antonucci(1993)]{ant93} Antonucci, R.\ 1993, \araa, 31, 473 
\bibitem[Bertram et al.(2007)]{ber07} Bertram, T., 
  Eckart, A., Fischer, S., et al.\ 2007, \aap, 470, 571 
\bibitem[Best \& Heckman(2012)]{bes12} Best, P.~N., \& Heckman, T.~M.\ 2012, \mnras, 2402 
\bibitem[B{\"o}ttcher \& Dermer(2002)]{bot02} B{\"o}ttcher, M., 
  \& Dermer, C.~D.\ 2002, \apj, 564, 86 
\bibitem[Bower et al.(2006)]{bow06} Bower, R.~G., Benson, 
  A.~J., Malbon, R., et al.\ 2006, \mnras, 370, 645 
\bibitem[Broderick et al.(2011)]{bro11} Broderick, A.~E., 
  Chang, P., \& Pfrommer, C.\ 2011, arXiv:1106.5494 
\bibitem[Cavaliere \& D'Elia(2002)]{cav02} Cavaliere, A., \& D'Elia, V.\ 2002, \apj, 571, 226
\bibitem[Chang et al.(2011)]{cha11} Chang, P., Broderick, A.~E., \& Pfrommer, C.\ 2011, arXiv:1106.5504 
\bibitem[Croton et al.(2006)]{cro06} Croton, D.~J., Springel, V.,
  White, S.~D.~M., et al.\ 2006, \mnras, 365, 11 
\bibitem[Danforth et al.(2010)]{dan10} Danforth, C.~W., Keeney, B.~A., Stocke, J.~T., Shull, J.~M., \& Yao, Y.\ 2010, \apj, 720, 976 
\bibitem[Dermer et al.(1992)]{dermer} Dermer, C. et al. 1992, A\&A, 256, L27 
\bibitem[Donoso et al.(2009)]{don09} Donoso, E., Best, P.~N., 
  \& Kauffmann, G.\ 2009, \mnras, 392, 617 
\bibitem[Evans et al.(2001)]{eva01} Evans, A.~S., Frayer, 
  D.~T., Surace, J.~A., \& Sanders, D.~B.\ 2001, \aj, 121, 1893 
\bibitem[Fabello et al.(2011)]{fab11} Fabello, S., Kauffmann, 
  G., Catinella, B., et al.\ 2011, \mnras, 416, 1739 
\bibitem[Feigelson \& Nelson(1985)]{fei85} Feigelson, E.~D., \& Nelson, P.~I.\ 1985, \apj, 293, 192 
\bibitem[Fossati et al.(1998)]{fos98} Fossati, G., Maraschi, 
  L., Celotti, A., Comastri, A., \& Ghisellini, G.\ 1998, \mnras, 299, 433 
\bibitem[Gaskell \& Sparke(1986)]{gas86} Gaskell, C.~M., \& Sparke, L.~S.\ 1986, \apj, 305, 175 
\bibitem[Ghisellini et al.(1998)]{ghi98} Ghisellini, G.,
  Celotti, A., Fossati, G., Maraschi, L., \& Comastri, A.\ 1998, \mnras, 301, 451 
\bibitem[Guillard et al.(2012)]{gui12} Guillard, P., Ogle, 
  P., Emonts, B., et al.\ 2012, arXiv:1201.1503 
\bibitem[Ho(2008)]{ho08} Ho, L.~C.\ 2008, \araa, 46, 475 
\bibitem[Kauffmann et al.(2012)]{kau12} Kauffmann, G., Li, 
  C., Fu, J., et al.\ 2012, arXiv:1202.2972 
\bibitem[Komatsu et al.(2011)]{kom11} Komatsu, E., Smith, 
  K.~M., Dunkley, J., et al.\ 2011, \apjs, 192, 18 
\bibitem[Laurent-Muehleisen et al.(1998)]{lau98} 
  Laurent-Muehleisen, S.~A., Kollgaard, R.~I., Ciardullo, R., et al.\ 1998, \apjs, 118, 127 
\bibitem[Lavalley et al.(1992)]{lav92} Lavalley, M.~P., 
  Isobe, T., \& Feigelson, E.~D.\ 1992, \baas, 24, 839 
\bibitem[Maraschi et al.(1992)]{maraschi} Maraschi, L. et al. 1992, \apjl, 397, L5 
\bibitem[Marscher et al.(2000)]{marscher} A.~Marscher \& R.~Protheroe 2000, AIPC, 515, 149
\bibitem[Neronov \& Semikoz(2009)]{ner09} Neronov, A., \& Semikoz, D.~V.\ 2009, \prd, 80, 123012 
\bibitem[Nesvadba et al.(2011)]{nes11} Nesvadba, N.~P.~H., Boulanger, F., Lehnert, M.~D., Guillard, P., \& Salome, P.\ 2011, \aap, 536, L5 \bibitem[Nilsson et al.(2003)]{nil03} 
  Nilsson, K., Pursimo, T., Heidt, J., et al.\ 2003, \aap, 400, 95 
\bibitem[Oca{\~n}a Flaquer et al.(2010)]{oca10} 
  Oca{\~n}a Flaquer, B., Leon, S., Combes, F., \& Lim, J.\ 2010, \aap, 518, A9 
\bibitem[Osterbrock \& Ferland(2006)]{ost06} Osterbrock, D.~E., \& Ferland, G.~J.\ 2006, Astrophysics of gaseous nebulae and active galactic nuclei, 2nd.~ed.~by D.E.~Osterbrock and G.J.~Ferland.~Sausalito, CA: University Science Books, 2006
\bibitem[Plotkin et al.(2012)]{plo12} Plotkin, R.~M., 
  Anderson, S.~F., Brandt, W.~N., et al.\ 2012, \apjl, 745, L27 
\bibitem[Riechers et al.(2011)]{rie11} Riechers, D.~A., 
  Carilli, C.~L., Maddalena, R.~J., et al.\ 2011, \apjl, 739, L32 
\bibitem[Sadler et al.(2007)]{sad07} Sadler, E.~M., Cannon, R.~D., 
  Mauch, T., et al.\ 2007, \mnras, 381, 211 
\bibitem[Saintonge et al.(2011)]{san11} Saintonge, A., 
  Kauffmann, G., Kramer, C., et al.\ 2011, \mnras, 415, 32 
\bibitem[Sbarufatti et al.(2005)]{sba05} Sbarufatti, B., 
  Treves, A., \& Falomo, R.\ 2005, \apj, 635, 173 
\bibitem[Scarpa et al.(2000)]{sca00} Scarpa, R., Urry, C.~M., 
  Falomo, R., Pesce, J.~E., \& Treves, A.\ 2000, \apj, 532, 740 
\bibitem[Schachter et al.(1993)]{sch93} Schachter, J.~F., 
  Stocke, J.~T., Perlman, E., et al.\ 1993, \apj, 412, 541 
\bibitem[Scoville et al.(2003)]{sco03} Scoville, N.~Z., 
  Frayer, D.~T., Schinnerer, E., \& Christopher, M.\ 2003, \apjl, 585, L105 
  \bibitem[Sikora et al.(1994)]{sikora} M.~Sikora et al.\ 1994, \apj, 421, 153
\bibitem[Skrutskie et al.(2006)]{skr06} Skrutskie, M.~F., 
  Cutri, R.~M., Stiening, R., et al.\ 2006, \aj, 131, 1163 
\bibitem[Smol{\v c}i{\'c} \& Riechers(2011)]{smo11} 
  Smol{\v c}i{\'c}, V., \& Riechers, D.~A.\ 2011, \apj, 730, 64 
\bibitem[Smol{\v c}i{\'c} et al.(2009)]{smo09} Smol{\v c}i{\'c}, V., Zamorani, G., Schinnerer, E., et al.\ 2009, \apj, 696, 24 
\bibitem[Stecker \& Scully(2010)]{ste10} Stecker, F.~W., \& Scully, S.~T.\ 2010, \apjl, 709, L124 
\bibitem[Trump et al.(2011)]{tru11} Trump, J.~R., Impey, C.~D., Kelly, B.~C., et al.\ 2011, \apj, 733, 60 
\bibitem[Urry et al.(2000)]{urr00} Urry, C.~M., Scarpa, R., O'Dowd, M., et al.\ 2000, \apj, 532, 816 
\bibitem[Willott et al.(2001)]{wil01} Willott, C.~J.,  Rawlings, S., Blundell, K.~M., Lacy, M., \& Eales, S.~A.\ 2001, \mnras, 322, 536 
\bibitem[Xia et al.(2012)]{xia12} Xia, X.~Y., Gao, Y., Hao, C.-N., et al.\ 2012, arXiv:1202.6490 
\bibitem[Young et al.(2011)]{you11} Young, L.~M., Bureau, M., 
  Davis, T.~A., et al.\ 2011, \mnras, 414, 940 
\end{thebibliography}
\end{document}